\documentclass[%
aip,
cha,
 amsmath,amssymb,
reprint,%
]{revtex4-1}

\usepackage{graphicx}
\usepackage{dcolumn}
\usepackage{bm}

\usepackage[utf8]{inputenc}
\usepackage[T1]{fontenc}
\usepackage{mathptmx}
\usepackage{etoolbox}
\usepackage{color,soul}

\makeatletter
\def\@email#1#2{%
 \endgroup
 \patchcmd{\titleblock@produce}
  {\frontmatter@RRAPformat}
  {\frontmatter@RRAPformat{\produce@RRAP{*#1\href{mailto:#2}{#2}}}\frontmatter@RRAPformat}
  {}{}
}%
\makeatother
\begin{document}

\preprint{AIP/123-QED}

\title[DH]{Influence of particle size polydispersity on dynamical heterogeneities in dense particle packings}
\author{Rajkumar Biswas}
\homepage{Rajkumar.Biswas@hhu.de}
\affiliation{ Soft Condensed Matter Group, Raman Research Institute, C. V. Raman Avenue, Sadashivanagar, Bangalore 560 080, INDIA
}

\author{Anoop Mutneja}
 \homepage{anoopmutneja@tifrh.res.in}
\affiliation{Tata Institute of Fundamental Research, 36/P, Gopanpally Village, Serilingampally Mandal, Ranga Reddy District, Hyderabad, 500107, India }%

\author{Smarajit Karmakar}
 \homepage{smarajit@tifrh.res.in}
\affiliation{Tata Institute of Fundamental Research, 36/P, Gopanpally Village, Serilingampally Mandal, Ranga Reddy District, Hyderabad, 500107, India }%

\author{Ranjini Bandyopadhyay}%
\homepage{ranjini@rri.res.in (corresponding author)}
\affiliation{ Soft Condensed Matter Group, Raman Research Institute, C. V. Raman Avenue, Sadashivanagar, Bangalore 560 080, INDIA
}%

\date{\today}

\begin{abstract}
The dynamics of dense particle packings near the jamming transition is characterized by correlated particle motion. The growth of dynamical heterogeneities, or strong spatial variations in the motion of the particles constituting the system, is a hallmark feature of slow glassy dynamics. We report here a systematic confocal microscopy study that characterizes the cooperative dynamics of fluorescently-labelled colloidal particles in dense aqueous suspensions. We demonstrate that jammed particulate suspensions can be fluidized by increasing the width of the particle size distribution. Our molecular dynamics simulations, performed to numerically investigate the effects of continuous-size polydispersity on dense particle packing dynamics,  show an excellent match with our experimental results. Besides shedding light on the fundamental aspects of particle-scale dynamics at the jamming-unjamming transition, our findings are significant in the processing of commonly-encountered dense suspensions such as paints, cosmetics, and food.
\end{abstract}
\maketitle
\section{Introduction}

If the temperature of a molecular system is lowered very rapidly, it can undergo kinetic arrest even while its structure remains disordered or liquid-like \cite{gotze1992relaxation, angell1995formation, berthier2011theoretical, karmakar2014growing}. The temperature at which this transition occurs, commonly referred to as the laboratory glass transition temperature ($T_G$), depends on the material properties and cooling rate. Changes in dielectric constant, viscosity, and thermodynamic properties like enthalpy, free volume, heat capacity, and thermal expansion coefficient are frequently used to identify the glass transition point \cite{stearns1966relationship, vcernovsek2002enthalpic, white2016polymer}. However, the structure and dynamics of molecular glass-forming materials, constituted by sub-nanometer sized entities, are inaccessible in laboratory experiments with standard measurement techniques. In this context, soft materials such as colloidal suspensions, constituted by macromolecules that statistically behave as scaled-up atoms, have been used extensively as models in the investigation of the physics of condensed matter systems \cite{marcelja1976phase, pusey2008colloidal}. Dense colloidal suspensions, for example, have been studied experimentally and through numerical simulations to unravel the structural and dynamical properties of glass-forming materials \cite{ganapathi2022structural}. As the sizes of the constituent particles in dense colloidal suspensions are several orders larger than those of atoms or molecules, their relaxation processes are amenable to measurement using common laboratory techniques such as bright field and confocal microscopy, dynamic light scattering, etc. \cite{pusey2008colloidal}. In contrast to a molecular system which can be transformed to a glass by a rapid temperature quench, the transition of a dense colloidal suspension to a glassy state can be achieved by increasing the volume fraction $\phi$ of the dispersed phase comprising colloidal particles \cite{pusey1987observation}. Particles constituting a dense system are trapped in cages formed by neighbouring particles and move in a correlated manner to break the caging \cite{weeks2002properties} and eventually attain an equilibrium state. A progressive slowing down of particle dynamics and a simultaneous growth of the characteristic length scale associated with cooperatively rearranging particles are ubiquitous near the glass transition volume fraction ${\phi_g}$ \cite{saha2014investigation, gadige2017study}.

The transition of a supercooled liquid to a glassy state is accompanied by a rapid rise in its relaxation time or viscosity \cite{berthier2005direct, karmakar2014growing, tamborini2015correlation, murarka2003diffusion}. Some theories of the glass transition \textit{e.g.}, random first order transition (RFOT) theory, predict a divergence of the relaxation time at a finite temperature or volume fraction. For colloidal systems, this theory suggests a glass transition at $\phi \approx 0.65$ in three dimensions \cite{adhikari2022dependence}, very close to the jamming transition which occurs at ${\phi}_{RCP} = 0.648$. Mode coupling theory (MCT), on the other hand, predicts a glass transition in these systems at a volume fraction $\approx 0.516$ for monodisperse systems \cite{gotze2009complex, janssen2018mode}. Interestingly, experiments show that particle dynamics persist above the MCT glass transition volume fraction, though they slow down considerably due to the growth of cooperatively rearranging regions (CRRs) \cite{brambilla2009probing}. 

It is not possible to prepare a size-monodisperse colloidal system in the laboratory, with size-polydispersity being an unavoidable part of any experimental system. For samples of negligible size polydispersity (smaller than 5\%), a dense system eventually settles in crystalline configurations. Since glasses are characterized by disorder and metastability that can be sustained only in the presence of particle size polydispersity \cite{auer2001suppression, hermes2010jamming}, the intrinsic polydispersity in colloidal systems has been exploited exhaustively for studying glassy dynamics.  It is well-known that the presence of size polydispersity shifts the glass transition volume fraction to a value that is higher than that predicted for a monodisperse system \cite{schaertl1994brownian}. An empirical formula has been reported where random close packing density can be calculated from polydispersity and skewness of the size distribution \cite{desmond2014influence}.   

Particle dynamics in dilute colloidal suspensions are purely diffusive, and particle displacements are well-characterized by Gaussian distribution functions \cite{sonntag1987diffusion}. While particle dynamics are restricted within cages for short observation times in dense particulate suspensions, a cage-hopping phenomenon, facilitated by the cooperative motions of particles, dominates at long waiting times \cite{hunter2012physics, kasper1998self}. Such particle displacements between cages results in a structural relaxation process characterized by an $\alpha$-relaxation time scale. The distribution of displacements of particles undergoing cooperative motion have been shown to deviate significantly from Gaussian statistics \cite{saltzman2006non,sengupta2014distribution}. The distribution of particle displacements is then characterized by clear exponential tails that become pronounced at low temperatures or high volume fractions. These tails are often associated with hopping motions by fast-moving particles and result in a multimodal and broad distribution of the diffusion constants in the system. Interestingly, the exponential tail in the probability distribution of particle displacements, often called the Van Hove function, can be completely rationalized by considering the growth of the length scale characterising the dynamical heterogeneity in the system \cite{bhowmik2018non}. Such distributions are often quantified using a non-Gaussian parameter $\alpha_2$ \cite{thorneywork2016gaussian}, defined as the ratio of the fourth to the second moments of the probability distribution of particle displacements. The time at which the peak in $\alpha_2$ appears, which is correlated but always smaller than the $\alpha$-relaxation time, is computed using the two-point overlap correlation function $Q(t)$ or the self intermediate scattering function $F_s(q,t)$, where $q$ is the wave vector and $t$ is the time \cite{dauchot2005dynamical, kob1995testing1}. The presence of peaks in $\alpha_2$, as reported in experimental studies \cite{bhowmik2018non, shell2005dynamic}, indicates spatiotemporal heterogeneities within the system.


Dynamical heterogeneities, which arise near the jamming transitions of granular materials, colloidal suspensions, molecular and supercooled liquids due to  very broad distributions of the dynamical relaxation rates, have been quantified successfully using four-point correlation functions \cite{dasgupta1992search, karmakar2009growing, karmakar2014growing, candelier2009building, maggi2012measurement, donati2002theory}. The dynamics of a dense, bi-disperse granular material under cyclic shear in a quasi two-dimensional geometry were found to be strongly correlated but spatially heterogeneous as its jamming transition was approached \cite{dauchot2005dynamical}. Using a novel method based on local topology \cite{abate2007topological}, the evolution of spatially heterogeneous dynamics and their connection with the local structure were studied for a quasi-two-dimensional granular system of air fluidized beads approaching the jamming transition. In another study involving a binary colloidal suspension in a quasi-two-dimensional geometry, the size of the dynamical heterogeneity was characterized in terms of a dynamical correlation length that grew rapidly during the crystal-to-glass transition \cite{yunker2010observation}.

Molecular dynamics simulations have reported that size polydispersity smears out the kinetic arrest of particles in dense packings due to a lubrication effect \cite{abraham2008energy,zaccarelli2015polydispersity}. In a recent study, Ludicina et al.\cite{PhysRevResearch.5.033121} investigated continuously polydisperse hard sphere fluids. Through a systematic analysis within the framework of MCT theory, they observed that polydispersity can significantly influence the glass transition point. Particularly, smaller particles are noted to play a crucial role in the structural relaxation of polydisperse systems. Heckendorf et al.\cite{PhysRevLett.119.048003} reported the impact of particle sizes near the tail of a Gaussian distribution. Through experimental investigation, they revealed that larger particles exhibit slower dynamics beyond a local volume fraction and that a clear link exists between local crowding and dynamical heterogeneity.Pednekar et al. \cite{10.1122/1.5011353} conducted a computer simulation study to investigate noncolloidal bidisperse and polydisperse systems. The rheological parameters extracted for both systems agreed well, which underscored the controlling influence of the maximum packing fraction in dense systems. Another theoretical study concluded that fragility of a supercooled liquid decreases with polydispersity due to significant alterations in the potential energy landscape \cite{abraham2008energy}. This was verified experimentally for soft colloidal glassy suspensions \cite{behera2017effects}. Often in the literature, the extent of heterogeneity is quantified in terms of the number of correlated particles, $N_{corr}$, which is estimated from the four-point dynamic susceptibility $\chi_4$ \cite{abate2007topological}. Theoretical studies have computed nonlinear susceptibilities to quantify the extent of cooperatively rearranging regions in spin glasses, dense colloidal systems, and Lennard Jones liquids \cite{bouchaud2005nonlinear, narumi2011spatial, glotzer2000time, lavcevic2003spatially}. $\chi_4$ exhibits a peak near the $\alpha$-relaxation time of the particles \cite{dasgupta1992search, karmakar2009growing, flenner2009subdiffusion}, with the height of the $\chi_4$ peak directly correlated with the average number of particles, $N_{corr}$, participating in cooperative rearrangements during structural relaxation. Although this gives a rough estimate of the number of particles that exhibit cooperative motion, a precise estimation of $N_{corr}$ requires one to compute the associated dynamical heterogeneity length scale, $\xi_D$, as a function of volume fraction or temperature $\textit{via}$ a detailed finite size scaling analysis of the peak height of $\chi_4(t)$ \cite{karmakar2009growing, chakrabarty2017block}, or from the wave vector dependence of the four-point structure factor, $S_4(q,t)$, computed at the $\alpha$-relaxation time \cite{flenner2010dynamic}. These analyses are significantly more challenging in experiments than in computer simulation studies due to experimental limitations in acquiring adequate statistics while probing dense systems.

We report systematic experiments and simulations to evaluate  the effect of continuous-size polydispersity on the correlated motion of particles in dense packings approaching kinetic arrest. We prepared dense colloidal microgel suspensions by synthesising fluorescent Poly(N-isopropylacrylamide) or PNIPAM particles with continuous-size polydispersities. We controlled the polydispersity index (PDI), defined as the ratio of the standard deviation to the mean of a Gaussian particle size distribution, by controlling reaction rates during particle synthesis. The evolution of particle dynamics with time was observed by tracking individual particles using a confocal microscope. We quantified the growth of dynamical heterogeneities in these polydisperse systems by calculating four-point dynamic susceptibility functions, $\chi_4(t)$ \cite{dasgupta1992search, lavcevic2003spatially, karmakar2009growing}, as a function of time $t$. Our measurements provide detailed information on the effect of particle polydispersity on the relaxation dynamics, spatiotemporal heterogeneity, and the number of correlated particles in cooperatively rearranging regions in disordered materials approaching their dynamically arrested states. Our work demonstrates that suppression of dynamic heterogeneities and enhanced suspension fluidization can be achieved by increasing particle PDI, medium temperature and suspension dilution. Our molecular dynamics simulations, performed to obtain $N_{corr}$ for crowded packings of particles having continuous-size polydispersity, match our experimental results very well.

\section{Material and methods}

\subsection{Functionalization of dye and synthesis of PNIPAM particles}

Behera et al. \cite{behera2017effects} reported a protocol for preparing continuously size-polydispersed particles. We synthesized polydisperse PNIPAM particles using the semi-batch method \cite{still2013synthesis, behera2017effects} in which the continuous size polydispersity (PDI) of the particles was controlled by varying the flow rate of the reaction ingredients during the synthesis process. The flow rate values for different batches are reported in Table.S1 in Supplementary Information. All chemicals for synthesis were purchased from Sigma-Aldrich. For our confocal experiments, we needed to attach fluorescein dye molecules to PNIPAM particles. To functionalize the dye, we added $3$ gm of fluorescein with $80$ ml of dry tetrahydrofuran in a $250$ ml round bottom flask. After mixing for $5$ min under nitrogen purging, $28$ ml of triethylamine was added. This mixture was stirred for 15 min in a nitrogen environment. The round bottom flask was kept in an ice bath, and $2$ ml of acryloyl chloride was slowly added under stirring conditions. The mixture was kept overnight to equilibrate and was eventually filtered to remove the triethylammonium chloride salt. The filtered sample was treated with nitrogen gas for $5$ min and then dried in a desiccator equipped with a rotor pump. The dried fluorescent crosslinker was stored at $4^{\circ}$C. 

To synthesize PNIPAM particles, we added $2$ gm N-isopropylacrylamide monomer (NIPAM $ \geq 99\%$), $0.05$ gm $N,N'$ – methylenbisacrylamide crosslinker (MBA), $0.0083$ gm of 2-aminoethylmethacrylate hydrochloride co-monomer (AEMA) and 0.1 gm of functionalized fluorescein in 50 mL Milli-Q water in a three-necked round bottom (RB) flask. The middle neck of the RB flask was connected to a reflux condenser. The RB flask was kept inside an oil bath to maintain a uniform temperature. The other two necks of the round bottom flask were used as $N_2$ inlet and outlet. The ingredients in the RB flask were stirred at 500 rpm for 20 minutes under $N_2$ purging. 30 mL of the sample was taken out using a syringe for later use. The temperature of the oil bath was increased to 80$^\circ$C, and 0.0114 gm ammonium persulfate (APS) dissolved in 2 mL of water was added to initiate polymerization  $via$ a free radical precipitation reaction. Nucleation sites started to form within 5 mins. The sample of 30 mL volume that was extracted earlier was next injected at different flow rates between 0.5 ml/min and 0.8 ml/min using a syringe pump to produce particles with distinct continuous size distributions that were quantified by calculating their PDI values from confocal images. After the process was completed, the latices were cooled down rapidly in an ice bath and were purified by repeated centrifugations. 

The size distributions of two representative batches of polydisperse fluorescent PNIPAM particles in suspension have been presented in Fig.1. PNIPAM particles are thermoresponsive (thermoresponsivity data acquired from an aqueous PNIPAM suspension prepared by us is presented in Fig.S1(a) of Supplementary Information) and show a swelling-deswelling transition at a lower critical solution temperature (LCST) of $\approx$ $32^\circ$ C \cite{heskins1968solution, hirokawa1984volume}.  The volume fraction of aqueous PNIPAM suspensions can therefore be tuned by controlling temperature across the LCST \cite{appel2016mechanics}. We measured the area fraction of the sample using a confocal microscope and compared the data across different area fractions. However, we could not measure the volume fraction of these particles using conventional methods because they deform under stress and their diameter changes with temperature.

\subsection{Confocal microscopy}

Below the LCST, PNIPAM particles swell by absorbing water and have a negligible refractive index mismatch with the surrounding water medium. It  is therefore impossible to distinguish individual PNIPAM particles in brightfield microscopy. As discussed in the previous section, we attached fluorescein dye to PNIPAM particles and imaged the fluorescent PNIPAM particles using a Leica TCS SP8 confocal microscope. The instrument has a resolution of approximately 210 nm in the XY plane. The dye attached to the particle was excited using a 488 nm Argon laser, and a photomultiplier tube was used  to record the output signal at a higher wavelength. We performed XY directional scanning to resolve the motion of individual particles in dense PNIPAM suspensions that were loaded in a sample cell constructed using a glass slide and a $\#1$ coverslip separated by two $\#0$ coverslips. It is generally advised to use \#1.5 coverslip for confocal microscope but we find the quality of the images were fairly good for the \#1 coverslip. The schematic of the sample cell is shown in Fig.S2 in Supplementary Information. After loading the sample, UV glue was used to seal the open ends of the sample cell that was kept inside a temperature-controlled enclosure box. Imaging was performed using a Nikon 100x oil immersion objective lens of numerical aperture, NA = 1.4. All experiments were performed in the temperature range 25$^\circ$C - 31$^\circ$C which lies below the lower critical solution temperature (LCST). The average sizes of the swollen particles do not change appreciably in this temperature range, as seen in Supplementary Information Fig.S1(b).

\begin{figure}[ht]
 \hskip -0.5in
\includegraphics[scale = 0.28]{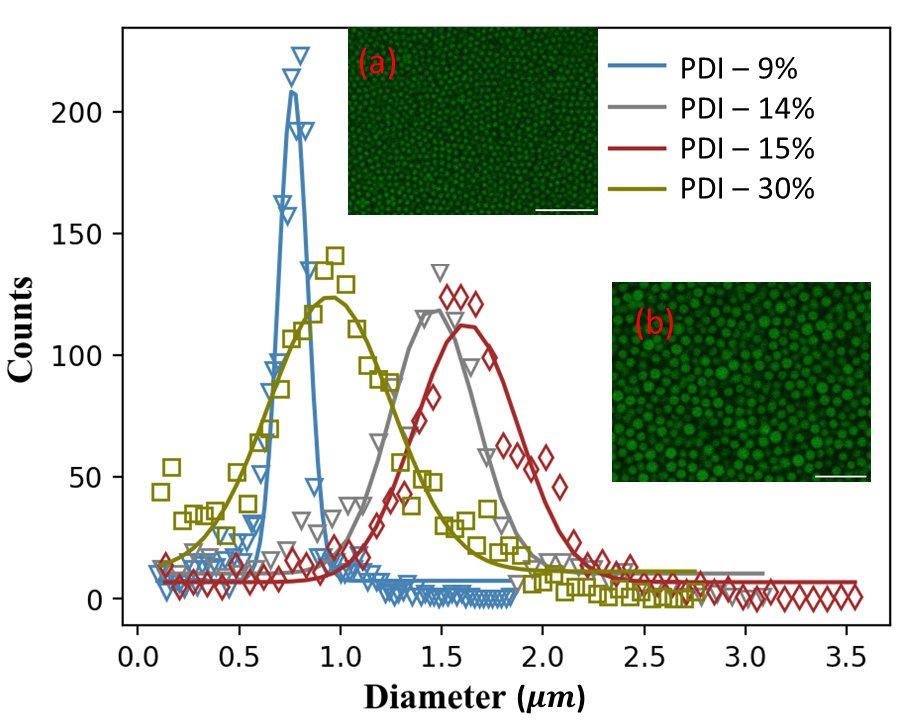} 
 \caption{Histogram plots of the diameter distributions of PNIPAM particles (hollow symbols), obtained from different synthesis batches. The solid curves are fits to Gaussian functions whose widths and heights are used to calculate the size polydispersity indices  (PDIs) of the PNIPAM particles. The insets show confocal images of dense suspensions having PDIs (a) 9\% and (b) 15\% with scale bar of 10 $\mu m$.}
 \label{fgr:PDI}
\end{figure}

Aqueous suspensions of fluorescent PNIPAM particles were loaded in the sample cell and imaged in two dimensions using a  confocal microscope. The depth of field of the setup was estimated to be $\approx 135$ nm. The imaging plane was fixed at $6-8$ $\mu m$ above the lower coverslip to avoid any wall effects. Images of different batches of PNIPAM particles in suspension, synthesized by varying the flow rates of the reaction ingredients, were analyzed to calculate particle size distributions and particle size polydispersity indices (PDI). For PDI calculation, we used a suspension of intermediate particle number density so that the edges of individual particles could be detected accurately. The confocal images were analysed using ImageJ \cite{schneider2012nih} to estimate the diameter distributions of the different batches of particles. The images were first changed to 8 bits and then a bandpass filter was applied. The out of focus particles from planes other than the plane of interest will contribute to the apparent polydispersity of the particles. The bandpass filter removed the out of focus particles to some extent which minimizes errors in polydispersity calculation. We next binarized the images to measure individual particle sizes. Error in determining mean size is expected to be roughly 3-4\%. The distributions of particle diameters (hollow symbols in Fig.\ref{fgr:PDI}) were fitted using Gaussian functions (solid lines in Fig.\ref{fgr:PDI}) and the mean $M$ and width $W$ of each distribution was computed. The fitted values can be found in Table.S2 in supplementary information. The size polydispersity indices (PDIs) of the particles synthesized in different batches by using different flow rates of the reaction ingredients were calculated as percentages using the formula, PDI $= 100*W/M$. 


\subsection{Estimation of mean square displacement (MSD)}
The mean square displacements (MSDs) of PNIPAM particles in quasi 2D experiments and Lennard Jones particles in 3D computer simulations were computed to estimate their average dynamics in samples of different particle area fractions $A_p$ (for our quasi 2D experiments), volume fractions $\phi$ (for the 3D computer simulations), particle PDIs, and medium temperatures using the formula:

\begin{equation}
<\Delta r^2 (t)> = \frac{1}{N} \sum_i <|\vec{r}_i (t_0 + t) -\vec{r}_i(t_0)|^2>_{t_0}   
\end{equation}

Here, $t$ represents the delay time, $t_0$ is the initial time of measurement of a particle trajectory, $\vec{r}_i$ is the position vector of the $i$-th particle, $N$ is the total number of particles ($N \approx 1000$), and $<>$ is an average over $t_0$.

\subsection{Estimations of four point susceptibility, $\chi_4(t)$, and number of correlated particles, $N_{corr}$}
We calculate two-point correlation functions ($Q_2$) and four-point susceptibilities (four-point correlation functions, $\chi_4$) to characterize the relaxation time scales and heterogeneous (correlated) dynamics in dense suspensions. We  define a two-point overlap correlation function $Q_2$ by averaging over a variable $Q_i$ over all particles and initial times $t_0$ \cite{dauchot2005dynamical}:

\begin{equation}
Q_i(a,t,t_0) = \exp(-\frac{\Delta r_i^2(t_0, t_0 + t)}{2a^2})
\end{equation}

\begin{equation}
Q_2 (a,t) = \frac{1}{N} \sum_i <Q_i (a,t,t_0)>_{t_0}
\label{Eq.Q2}
\end{equation}

where $\Delta r_i(t_0, t_0 + t)$ is the displacement of the $i$-th particle during a delay time $t$ (estimated from the initial time $t_0$), $a$ is a probe length that removes the decorrelations that might happen due to vibration of particles in cages in a dense suspension and $N$ is the total number of particles. In Eqn.2, the exponential form arises from the calculation of a spatial overlap function \cite{dauchot2005dynamical}. If the displacement $\Delta r_i(t_0, t_0 + t)$ of the $i$-th particle is less than the probe length $a$, $Q_i$ (Eqn.2) will be close to 1,  while for displacements larger than $a$, $Q_i$ will be closer to 0. We have systematically varied the probe length, $a$, across a range of length scales. Fluctuations in $Q_2$ were subsequently quantified by estimating the four-point susceptibility or the four-point correlation function $\chi_4$ as follows:

\begin{eqnarray}
\chi_4 (a,t) = N( \frac{1}{N} \sum_i <Q_i(a,t,t_0)^2>_{t_0} - \nonumber\\ \frac{1}{N} \sum_i <Q_i(a,t,t_0)>_{t_0}^2) 
\label{Eq.X4}
\end{eqnarray}

Finally, we identified the length scale that corresponds to the strongest peak in $\chi_4$, and therefore maximally non-Gaussian dynamics. This length scale is  designated as the optimal probe length, $a^*$. We have used $a^*$ to calculate the two-point correlation function using Eqn.3, and eventually the four-point correlation function using Eqn.4 to quantify the extents of dynamically heterogeneous regions in the sample. The number of correlated particles in a dynamical heterogeneity, $N_{corr}$, was estimated using the formula \cite{abate2007topological}:
\begin{equation}
N_{corr} = \frac{\chi_4^*}{1 - Q_2^*}
\label{Eq.NCorr}
\end{equation}

where $\chi_4^*$ is the peak value of $\chi_4(t)$ at the optimal probe length $a^*$ and $Q_2^*$ is the magnitude of $Q_{2}(t)$  at the delay time corresponding to the peak in $\chi_4$. All analyses and numerical calculations were performed using Python 3.10 (Jupyter notebook).

\section{Results and Discussion}

\subsection{Experimental Results}

\begin{figure*}

  \includegraphics[height=6.0cm]{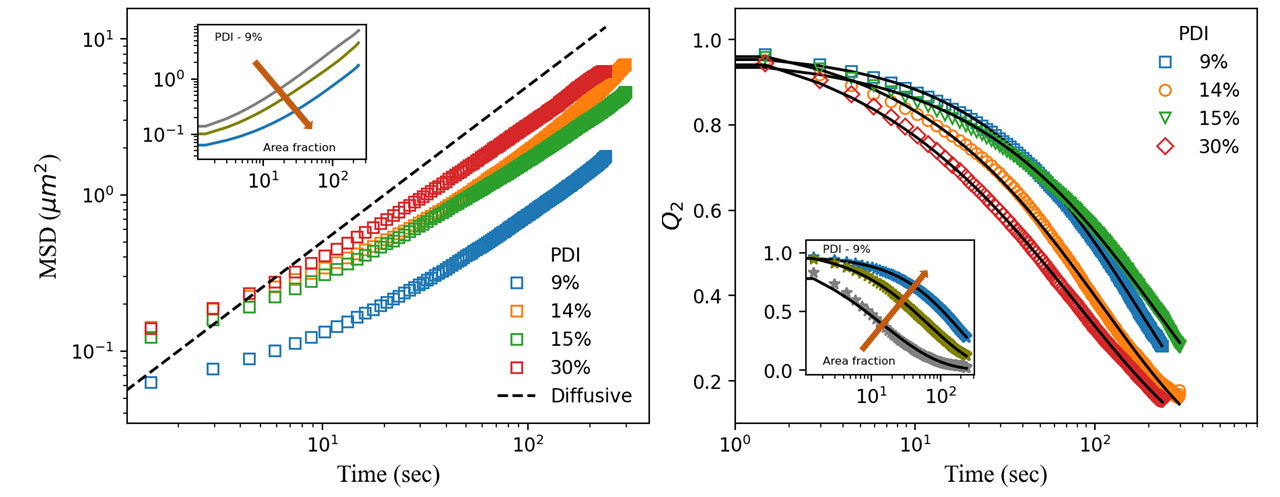}
  \caption{(a) Mean square displacements (MSDs) $vs.$ time for dense PNIPAM suspensions constituted by particles of different PDIs at a particle area fraction $A_p\approx 0.65$. The black dashed line represents diffusive dynamics. In the inset, MSDs are plotted for PNIPAM suspensions constituted by particles of PDI = 9\% for three different area fractions $A_p$ (0.52, 0.59, 0.65), with the arrow pointing towards the direction of increasing area fraction. (b) Two-point correlation functions, $Q_2$, at $a^{*}$ (Supplementary Table S3) are plotted $vs.$ time, for suspensions of different particle PDIs at a fixed area fraction ($A_p = 0.65$). The black solid curves are fits to stretched exponential functions. $Q_2$ plots for PNIPAM suspensions of different area fractions $A_p$ (0.52, 0.59, 0.65) for a fixed particle PDI = 9\% are shown in the inset. The arrow indicates the direction of increasing area fraction.}
  \label{fgr:msd}
\end{figure*}

To compute the particle dynamics, image sequences were acquired for $3-4$ mins at $0.679$ frames/sec, and $1000$ PNIPAM particles from the central regions of the images were tracked using a video spot tracker software (Computer Integrated Systems for Microscopy and Manipulation) \cite{tracker}. The pixel length of the images is approximately 80 nanometers. The positional uncertainty of the tracking algorithm is on the order of sub-pixel length. If a particle moved beyond the plane of interest during the experiment, tracking of that particle was discontinued. We have calculated the area fractions, $A_p$, of PNIPAM particles in dense  suspensions by binarizing two-dimensional images after setting appropriate intensity threshold values such that individual particles could be distinguished from the background. Multiple measurements of the same sample were performed at different times to detect the presence of aging. We did not find significant change in MSD curves measured at different times. MSDs were determined $vs.$ delay time $t$ and plotted in Fig.\ref{fgr:msd}(a) for aqueous suspensions of PNIPAM particles of four distinct PDI values, prepared at a high particle area fraction ($A_p \approx 0.65$). MSD data for particles having a fixed PDI ($=9\%$) and of varying suspension area fractions, $A_p$, are shown in the inset. The observed enhancement in MSDs with increasing PDI indicates the onset of unjamming or fluidization, and is reminiscent of earlier results \cite{PhysRevResearch.5.033121,zaccarelli2015polydispersity,behera2017effects}. Furthermore, we see from the inset of Fig.\ref{fgr:msd}(a) that the calculated MSDs decrease with  increase in particle area fraction due to constrained particle motion in increasingly jammed environments. Increasing suspension area fraction and decreasing particle PDI, therefore, have analogous effects on suspension jamming. In Fig.\ref{fgr:msd}(a), we note the emergence of very weak plateaus at low delay times in the MSDs of suspensions of particles having lower particle PDI values due to constrained motion \cite{hunter2012physics,weeks2002subdiffusion}. The MSD increases at later times and approaches a line of slope 1 ($\Delta r^2(t) \propto t$ marked by a dashed line), which indicates onset of the cage escape process $via$ diffusion \cite{kegel2000direct}. We note that the extent of the weak plateau in the MSD reduces with increase in PDI and decrease in area fraction, thereby pointing to enhanced fluidization of the system under these conditions.

\begin{table*}
\caption{\label{tab:table3} Relaxation time scales, $\tau$, and stretching exponents, $\beta$, extracted from fits of stretched exponential functions to two-point correlation functions, $Q_2$, estimated from both experiments (black solid lines in Fig.2(b)) and simulations  (black solid lines in the inset of Fig.4(b)). The peak height values of $\chi_4$, $\chi_4^*$, extracted from experiments and simulations (Figs.3 and 4(b)).}
\vskip +0.1in
\begin{ruledtabular}
\begin{tabular}{ccccc|ccccc}
 &\multicolumn{2}{c}{$Experimental$} &\multicolumn{7}{c}{$Simulation$}\\
 PDI & $A_p$ & $\tau$ (sec) & $\beta$ & $\chi_4^*$ & PDI & $\phi$ & $\tau$ (steps) & $\beta$ & $\chi_4^*$ \\   \hline

9\% & 0.65 & 181.8$\pm$0.5 & 0.80 & 4.06 & 13.06\% & 0.623 & 2211.7$\pm$16.0 & 0.70 & 4.35\\
14\% & 0.66 & 111.4$\pm$1.0& 0.67 & 1.81 & 17.37\% & 0.623 & 790.8$\pm$6.8 & 0.59 & 2.26\\
15\% & 0.65 & 229.5$\pm$1.1& 0.70 & 1.80 & 21.77\% & 0.623 & 229.8$\pm$4.3& 0.58  & 1.33\\
30\% & 0.65 & 80.4$\pm$0.4 & 0.60& 0.82 & 38.67\% & 0.623 & 156.6$\pm$3.1 & 0.55 & 0.41
 
\end{tabular}
\end{ruledtabular}
\end{table*}

The two-point and four-point correlation functions, computed using Eqns.3 and 4, for data acquired from a dense sample with $A_p = 0.65$ and PDI = $9\%$ are displayed in Supplementary Fig.S3 for different probe lengths $a$. As expected, $Q_2$ relaxes at a faster rate and $\chi_4$ peaks are shorter when probe lengths $a$ are sub-optimal. As $a$ increases, the decay time of the two-point correlation function becomes slower, while the peak height of $\chi_4$ grows higher. When $a$ exceeds an optimal value $a^*$, the peak in $\chi_4$ is seen to reduce. The optimal probe lengths $a^*$ at which the fluctuations of $Q_2(t)$, or the peak heights of $\chi_4(t)$, are maximum for all the samples are tabulated in Table.S2 of Supplementary Information. The sample dynamics are expected to be maximally heterogeneous at $a^*$ \cite{dauchot2005dynamical, abate2007topological}.



\begin{figure}
\centering
  \includegraphics[scale=0.7]{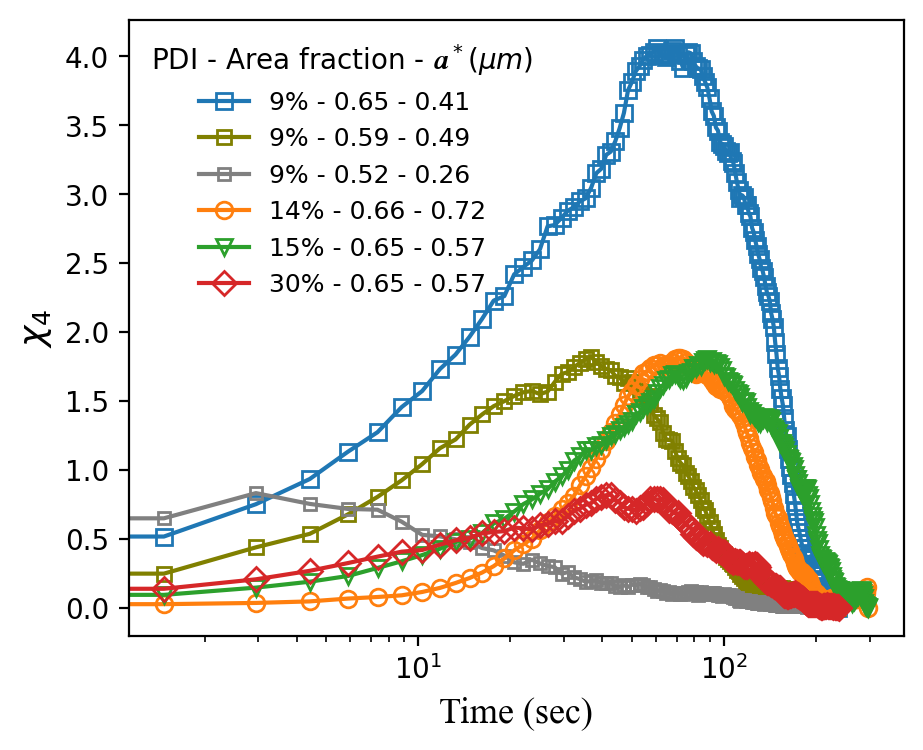}
  \caption{$\chi_4$ plotted $vs.$ time for probe lengths $a^*$ for dense PNIPAM suspensions of different size polydispersities (PDIs) and area fractions $A_p$.}
  \label{fgr:X4}
\end{figure}

In Fig.2(b), we display the time-decays of normalized $Q_2$ estimated from the displacement squares (Fig.2(a)) acquired from particle trajectories. Normalization is done by dividing all the values with the first data point of the data set to make the initial value 1. We see that $Q_2$ decays faster with increase in particle PDI. For a fixed low particle PDI of $9\%$, $Q_2$ relaxes faster with decreasing area fraction $A_p$ (inset of Fig.2(b)). We have fitted the two-point correlation data with stretched exponential functions, $Q_2 = A \exp(-(t/\tau)^\beta)$ \cite{diaz2020microscopic} to compute the characteristic decay times $\tau$. In the expression for $Q_2$, the fitting parameters $A$ and $\beta$ are the prefactor and the stretching exponent respectively where fitted values of $A$ are close to 1. The other fitted parameters $\tau$ and $\beta$ are displayed in Table I. We note from the values of $\tau$ that the relaxation process speeds up considerably with increase in particle PDI. We note that while fits to experimental data reveal the actual decay times,  the time scales extracted from simulations (to be discussed later) are related to the number of simulation steps. We would like to highlight that our fitted $\beta$ values roughly match with another recent investigation of polydisperse systems by Laudicina et. al \cite{PhysRevResearch.5.033121}. As discussed earlier, the four-point susceptibility, $\chi_4$, can be physically interpreted as a measure of the spatiotemporal variation of $Q_2$. In Fig.\ref{fgr:X4}, $\chi_4(t)$ extracted from experiments with suspensions of different PDIs and area fractions, $A_p$, are plotted at the optimal probe lengths $a^*$ evaluated as described earlier and tabulated in Table.S3 in Supplementary Information. Finally, we have tabulated the peak heights $\chi_4^*$ of the four-point susceptibility functions, $\chi_4$, in Table I to systematically study the extent of spatiotemporal heterogeneities in dense PNIPAM suspensions of different area fractions constituted by particles of different PDIs.

It has been reported that while the peak height of $\chi_4$ is proportional to the extent of dynamical heterogeneity present in the system, the time scale corresponding to the peak position of $\chi_4(t)$ is comparable to the $\alpha$-relaxation time scale of the sample \cite{karmakar2016overview}. From the data plotted in Fig.\ref{fgr:X4}, we see that the $\chi_4(t)$ peak becomes weaker as area fraction of the suspension, $A_p$, is decreased and as particle PDI is increased. A decrease in $\chi_4$ peak indicates a loss of spatiotemporal correlations. The observed fluidization of the sample with increasing dilution, achieved by decreasing suspension area fraction, is consistent with the predictions of the jamming phase diagram \cite{liu1998jamming}. We have also checked that the peak value of $\chi_4$ decreases with increasing temperature for fixed PDI and area fraction. The data is plotted in Fig.S4 in  Supplementary Information. When the sample temperature is raised or particle number density is reduced, the accelerated diffusion of PNIPAM particles unjams the suspension. Also as already discussed, the data in Fig.\ref{fgr:X4} clearly demonstrates that an increase in PDI enhances the lubrication effect as described in earlier reports \cite{abraham2008energy, zaccarelli2015polydispersity}, thereby resulting in suspension unjamming.

\subsection{Molecular dynamics simulation}

We next verify our experimental results by numerically studying the effects of continuous particle size polydispersity on the dynamics of particle packings. We model inter-particle interactions using a Lennard Jones potential, $V_{LJ}(r) = 4\epsilon[(\sigma / r)^{12} - (\sigma / r)^6]$ for $r \leq L_{cuttoff}$, $V_{LJ}(r) = 0$ for $r > L_{cuttoff}$, where $\epsilon$ is the dispersion energy and $\sigma$ corresponds to the minimum center-center distance between particles at which the potential is zero, $L_{cuttoff}$ is $2^{1/6}\sigma_{ij}$, where $i$ and $j$ denote two different particles and $\sigma_{ij} = (\sigma_{i} + \sigma_{j})/2$. The Lennard-Jones potential has been used extensively to study supercooled colloidal suspensions and their approach towards the glass transition. In our attempt to numerically study the effect of particle size polydispersity on the cooperative particle dynamics and dynamical heterogeneity in the sample, we have selected the Lennard Jones potential \cite{kob1995testing} because of its ubiquity in the study of glassy dynamics. In our experiments, samples were in a 3D cell but image acquisition was performed in quasi 2D. The acquired images are restricted to the XY plane with a small field of depth. Technical difficulties in quantifying experiments with dense suspensions, for example, in the determination of an absolute volume fraction and in avoiding systematic errors during particle tracking, motivated us to verify our central experimental results {\it{via}} 3D simulations. We note that the key results of our study are expected to be insensitive to details of the inter-particle potential or number of spatial dimensions. 

We have introduced polydispersity in size by incorporating a normally distributed function having mean = 0 and standard deviation = 1. The lower ends of the distribution have been removed to exclude particles with negative sizes. The generated normal distribution (ND) is truncated at -2 and +2 (this truncation is not performed for experimentally calculated polydispersity index values, as experimentally, diameters cannot have negative values). In this work, the distribution of sizes ($\sigma_i$) of $50000$ particles is defined as $\sigma_i = 1 + \frac{PDI}{100} * ND$, where $ND$ is the truncated normal distribution function. Particles generated using a stochastic process can introduce significant error due to diameter disorder at low temperatures. \cite{PhysRevE.106.064103}

\begin{figure*}[ht]
 \centering
 \includegraphics[height=6.0cm]{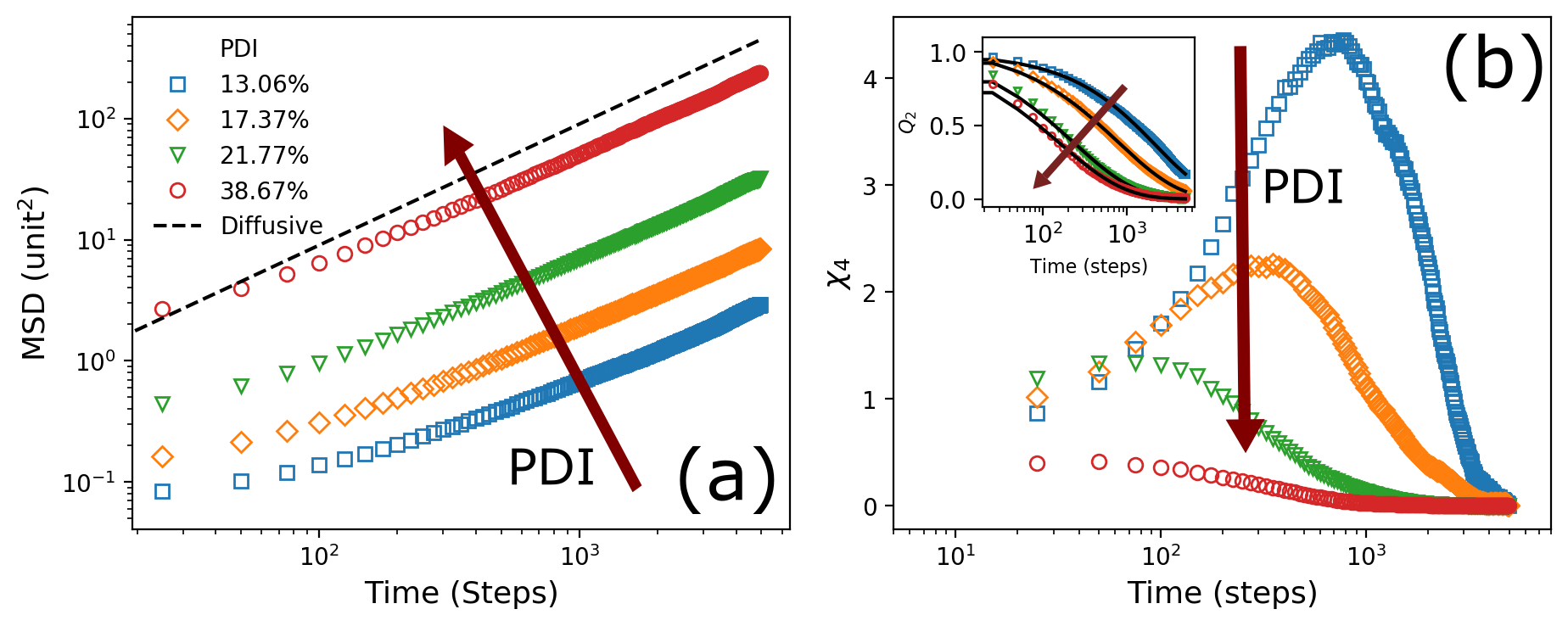}
 \caption{(a) Mean square displacements, MSDs, of the simulated particles $vs.$ time, shown for different particle PDI values for volume fraction $\phi$ = 0.6230. (b) Four-point susceptibilities, $\chi_4$, $vs.$ time for the same PDI and $\phi$ values. In the inset, the corresponding two-point correlation functions, $Q_2$, are shown. The correlation functions are calculated at $a=a^{*}$. The arrows point towards the direction of increasing particle PDI.}
 \label{fgr:MSD_Sim}
\end{figure*}
In the molecular dynamics (MD) simulations, the systems were initialized at high temperatures to avoid artefacts arising from the initial particle configuration for 5000 steps. Furthermore, the equilibration steps were performed in canonical ensemble (NVT) using Nose Hoover thermostat followed by simulation runs in microcanocical ensemble (NVE) by fixing the volume fraction, $\phi$, at $0.6230$ $\pm$ $0.0005$ for a system with $N = 50000$ particles and for particle PDI values $13.06\%$, $17.37\%$, $21.77\%$ and $38.67\%$. Volume fraction has been calculated from the total volume occupied by the particles using their diameters ($\sigma_i$) and the total box volume. Generally number density is used to study LJ systems but as we have changed PDI here, it is more accurate to use volume fraction for comparison. First, the system is mixed at a high temperature T = 1.5 for 5000 steps and then it is equilibrated for 100000 steps. The equilibration is performed much longer than the relaxation timescales of these systems to eliminate aging effects that might influence the dynamics of the system. After the equlibration is complete, we have recorded the trajectories. Time steps used for all the simulation is 0.0025. The temperature $T$ is fixed at 0.9 (Lennard-Jones unit). We selected seven different regions from the system, each having $1000$ particles. These parameters were chosen to ensure that the simulations match the experimental conditions. The usual metrics, the MSDs, $Q_2(t)$s and $\chi_4(t)$s are calculated by averaging over all seven regions to ensure adequate statistics and maximum overlap with experimental conditions. In Fig.\ref{fgr:MSD_Sim}(a), the simulated mean square displacements (MSDs) are plotted for dense suspensions comprising particles with different PDIs. We see that the computed MSD values are larger for samples with higher particle PDIs, thereby verifying our experimental result that increasing continuous size polydispersity can drive the fluidization of jammed particulate suspensions.  

Next, four-point correlation functions  corresponding to the optimal probe lengths, $a^*$, extracted from simulations using Eqn. 4 and tabulated in Table.S3 of Supplementary Information, are plotted $vs.$ time in Fig.\ref{fgr:MSD_Sim}(b). As in the experiments, $a^*$ is chosen independently for each sample and represents the value of $a$ at which the peak height of $\chi_4(t)$ is maximum (Table.S3 in Supplementary Information). We note that $\chi_4^*$, the height of the $\chi_4(t)$ peak, reduces and shifts to an earlier time for a particle assembly that is characterized by a higher PDI. These results are also consistent with our experimental observations. We therefore note that heterogeneous dynamics driven by cooperative particle motion are most significant in dense suspensions characterized by lower particle PDIs. Two-point correlation functions, $Q_2$, for optimal probe length values $a^*$ and several particle PDIs are plotted in the inset of Fig.\ref{fgr:MSD_Sim}(b). In a result that verifies our earlier experimental observation plotted in Fig.2(b), we note that the two-point correlation functions, $Q_2$,  decay faster as PDI is increased at a fixed area fraction. The relaxation time scales indicate that the systems are in the mildly supercooled state.

Finally, we have performed simulations while also increasing temperatures ($T$ = $0.8, 0.9$ and $1.0$) and volume fractions ($\phi$ = $0.617$, $0.620$ and $0.623$) for particle packings having the lowest PDI ($13.06\%$). This data, presented in Supplementary Information Figs.S5 (a-d), again confirms our earlier experimental results that particle dynamics accelerate with increasing temperature (Fig.S5(a)) and decreasing volume fraction (Fig.S5(b)). The gradual increase in the slope of the MSD plot with increasing temperature in Fig.S5(a) and decreasing $\phi$ in Fig.S5(b) point to enhanced diffusion under these conditions. We find that $\chi_4^*$ decreases with increasing temperature and decreasing volume fraction (increasing dilution), which suggests reduction in the extent of dynamical heterogeneities under these conditions (Supplementary Information Figs.S5 (c) and (d)).

\subsection{Estimation of $N_{corr}$ and a modified jamming-unjamming phase diagram}

\begin{figure}[ht]
\centering
  \includegraphics[height=6cm]{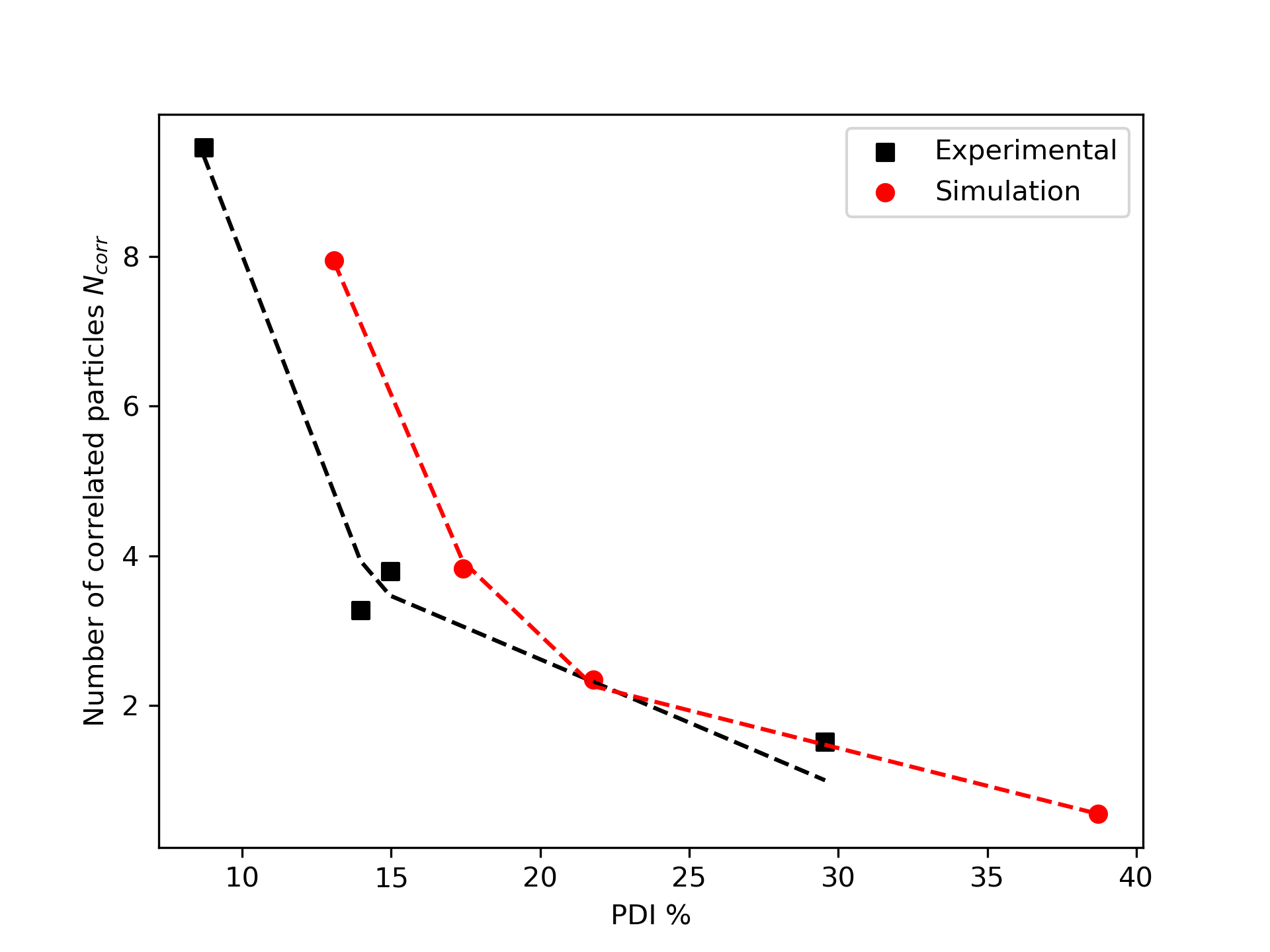}
  \caption{Number of correlated particles in cooperatively rearranging regions due to dynamical heterogeneity, $N_{corr}$, obtained from experiments (black) and simulations (red), for dense suspensions constituted by particles having different PDIs.}
  \label{fgr:NCorr}
\end{figure}

We estimate the characteristic length scales of the cooperatively rearranging regions in dense polydisperse suspensions by plotting  $N_{corr}$,  computed from both experiments and simulations using Eqn. 5, vs. PDI, in Fig.\ref{fgr:NCorr}. In spite of the differences in interparticle potential and system dimensionality in our experiments and simulations, we note that $N_{corr}$ decreases with increasing PDI in both cases. We conclude therefore that dense particulate systems can be unjammed by increasing the continuous size polydispersity of the constituent particles at constant medium temperature and particle density.

\begin{figure}[ht]

  \includegraphics[height=7cm]{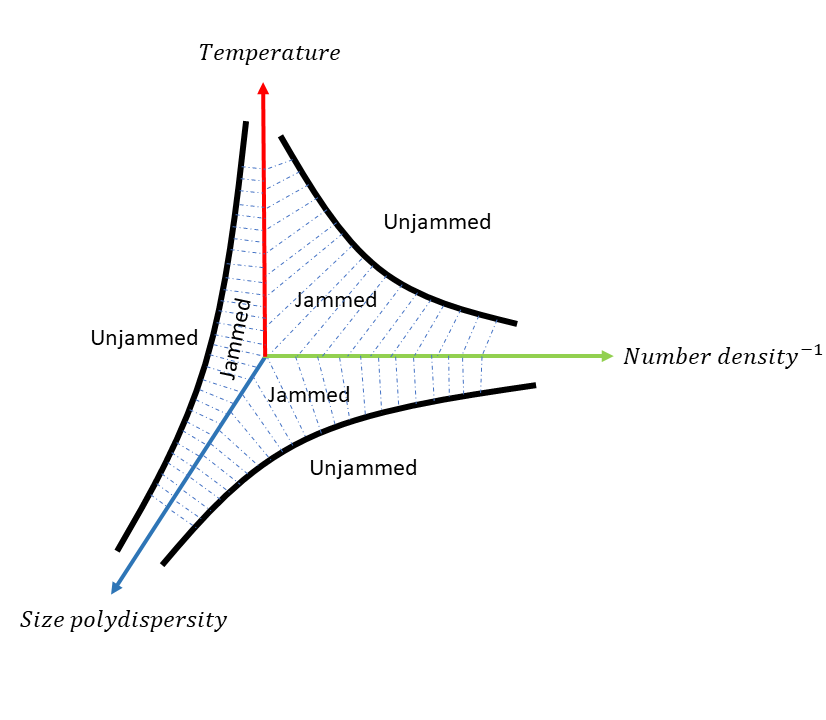}
  \caption{ A three dimensional jamming-unjamming phase diagram based on our experiments and simuations, with particle size polydipersity (PDI), inverse of particle number density (related to area fraction in experiments and volume fraction in simulations) and temperature as the 3 axes. The jammed region (blue dashed lines) lies around the origin. Arbitrary black solid lines are drawn to visualize the jamming to unjamming transition, achieved by increasing temperature and size polydispersity, or by decreasing particle number density. }
  \label{fgr:phaseDia}
\end{figure}

It is well known that the random close packing fraction of a suspension shifts towards a higher volume fraction with increasing size polydispersity \cite{schaertl1994brownian}. As the smaller particles in polydisperse particle packings are more mobile, they effectively fill the voids between larger particles to create new packing configurations that are more compact than those achieved in size-monodisperse particle assemblies. Consequently, the dynamics of polydisperse particle assemblies can remain fluid-like above the random close packing volume fraction for monodisperse systems \cite{hermes2010jamming}. By performing systematic experiments and simulations, we have demonstrated here that jammed suspensions can be fluidized effectively by increasing the particle polydispersity index (PDI) and sample temperature (T) and by decreasing particle density / concentration ($A_p$ in experiments and $\phi$ in simulations). As already discussed for particle packing with large PDIs, the small particles having higher mobilities can effectively enhance the lubrication effect \cite{abraham2008energy}. Furthermore, increasing temperature and dilution also contribute to increasing particle mobility, thereby accelerating the average dynamics of the suspension. Our experiments and simulations therefore provide an intuitive description of jamming-unjamming dynamics by using experimental and simulational results consisting particles with different inter-particle interactions and dynamical analyses for different dimensions.

Finally, we propose a phase diagram to highlight the jamming to unjamming transition in dense particle packings. While fluidization by increase in temperature and applied stress, or {\it via} decrease in packing fraction and attractive interactions  have been discussed in earlier works \cite{liu1998jamming, trappe2001jamming, farris1968prediction, gadige2017study}, our study suggests a new strategy of suspension unjamming \textit{via} an increase in particle PDI. A change in PDI shifts the $\phi_{RCP}$ of the system to higher values. The extent of  dynamical heterogeneity decreases under these consitions, leading to system fluidisation and significant changes in the emergent dynamical properties. Our three-dimensional jamming-unjamming phase diagram, displayed in Fig.\ref{fgr:phaseDia}, highlights that enhancement of particle polydispersity is an alternate new route to achieve unjamming.

\section{Conclusions}
Using particle tracking experiments and molecular dynamics simulations, we report the presence of dynamically heterogeneous regions in dense particle packings for low particle polydispersity indices (PDIs) and at low temperatures. We have presented here the first direct measurements of the dependence of the size of dynamical heterogeneity (in terms of $N_{corr}$) on particle polydispersity index (PDI). In contrast to earlier work in which two batches of monodisperse particles of incommensurate sizes were mixed to prepare a bidisperse glass \cite{candelier2009building, zhang2011cooperative, avila2014strong}, the present work estimates the dynamics of jammed states constituted by particles having continuous size distributions. We compute mean square displacements to estimate the average dynamics of particles in dense packings and extract multipoint correlation functions to characterize their heterogeneous dynamics. The weakening of particle caging with increase in PDI is evident from our analyses. The observed speeding up of the structural relaxation process with increasing particle PDI can be explained by considering the presence of smaller particles in dense systems facilitates larger movement by creating space using efficient void-filling. The particle dynamics are expected to be maximally heterogeneous at the time scale of the structural relaxation process due to the large non-Gaussian displacements of a small fraction of constituent particles during diffusion \cite{saltzman2006non, shell2005dynamic}. For dense colloidal suspensions, therefore, particle diffusion is strongly dependent on the particle size distribution, which dictates the characteristic time scales associated with particle dynamics in crowded environments.

The spatiotemporal heterogeneities in the dynamics of dense suspensions are studied here by changing the particle area fraction $A_p$ (in experiments) and volume fraction $\phi$ (in simulations) at a fixed PDI, and by also changing particle PDIs at a fixed $A_p$ or $\phi$. The time and length scales associated with spatiotemporal heterogeneities are extracted from four-point susceptibilites, $\chi_4$. The growth of correlations is more prominent for suspensions constituted by particles having narrower size distributions. The presence of heterogeneous and correlated dynamics in dense suspensions, as shown by us \textit{via} careful experiments and molecular dynamics simulations, points to the analogous natures of the jamming and glass transitions. Despite the differences in our experimental and simulation protocols, the key results from our experiments and simulations exhibit a notable similarity, with the dynamical heterogeneity decreasing with particle PDI in both cases. An increase in the number of particles undergoing correlated motion in suspensions characterized by lower PDIs indicates the existence of a growing length scale \cite{tah2021understanding} during the kinetic arrest phenomenon. Besides demonstrating the disappearance of correlated and heterogeneous particle dynamics with increasing particle size distributions, we also show that dynamical heterogeneity vanishes with increase in dilution and temperature. Our work clearly illustrates that the kinetic arrest phenomenon can be controlled by changing particle size polydispersities. In future studies, particles labelled with different dyes can be used for size-resolved analyses of local dynamics. It would be interesting to study the influence of local order on the emergence and disappearance of dynamical heterogeneities in future work. 

\section*{Supplementary material }

The supplementary material provides information on the synthesis process and characterization of PNIPAM particles. Figures illustrating the dependence of correlation functions on probe length are included, and a table presents our estimates of the optimal probe lengths. Additionally, the supplementary information file contains figures that provide further insights into temperature and volume fraction dependencies of the computed correlation functions.

\section*{Author Declarations}

\subsection*{Conflict of Interest}
The authors have no conflicts to disclose.
\subsection*{Contribution statement}
\textbf{Rajkumar Biswas}: Data curation; Formal Analysis; methodology; software; visualization; original draft preparation.
\textbf{Anoop Mutneja}: Methodology; software.  
\textbf{Smarajit Karmakar}: Supervision; draft review and editing; resources.
\textbf{Ranjini Bandyopadhyay}: Conceptualization; supervision; project administration; original draft preparation; draft reviewing and editing; resources.

\section*{Acknowledgements}
The authors thank Yatheendran K.M. and Vasudha K.N. for their help during the experiments. R. Biswas and R. Bandyopadhyay would like to acknowledge Lucas Goehring for useful discussion and Sayantan Chanda for his suggestions. We thank CISMM at UNC-CH, supported by the NIH NIBIB (NIH 5-P41-RR02170) for tracking software. S. Karmakar would like to acknowledge support through the Swarna Jayanti Fellowship Grant No. DST/SJF/PSA-01/2018-19 and SB/SFJ/2019-20/05.

\section*{Data Availability Statement}

The data that support the findings of this study are available from the corresponding author upon reasonable request.

\section*{References}
\bibliography{sorsamp}

\end{document}


\renewcommand{\figurename}{Supplementary Fig.}
	\setcounter{table}{0}
	\renewcommand{\thetable}{S\arabic{table}}%
	\renewcommand{\tablename}{Supplementary Table}
	\setcounter{figure}{0}
	\makeatletter 
	\renewcommand{\figurename}{Supplementary Fig.}
	\setcounter{figure}{0}
	\makeatletter 
	\renewcommand{\thefigure}{S\arabic{figure}}
	\setcounter{section}{0}
	\renewcommand{\thesection}{ST\arabic{section}}
	\setcounter{equation}{0}
	\renewcommand{\theequation}{S\arabic{equation}}
 
	\title{\color{black}\textbf{\underline{Supplementary Information}\\ Influence of particle size polydispersity on dynamical heterogeneities in dense particle packings}}

	\author[1, $\dagger$]{Rajkumar Biswas}
	\affil[1]{\textit{Soft Condensed Matter Group, Raman Research Institute, C. V. Raman Avenue, Sadashivanagar, Bangalore 560 080, INDIA}}
	\author[2, $\ddagger$]{Anoop Mutneja}
        \author[2, $\ddag$]{Smarajit Karmakar}
 \affil[2]{\textit{Tata Institute of Fundamental Research, 36/P, Gopanpally Village,
Serilingampally Mandal, Ranga Reddy District, Hyderabad, 500107, India}}
	\author[1,*]{Ranjini Bandyopadhyay}
	\date{\today}
	
	\footnotetext[2]{Rajkumar.Biswas@hhu.de}
	\footnotetext[3]{anoopmutneja@tifrh.res.in}
        \footnotetext[4]{smarajit@tifrh.res.in}
	\footnotetext[1]{Corresponding Author: Ranjini Bandyopadhyay; Email: ranjini@rri.res.in}
	\maketitle
	\pagebreak
        

                    \section{Flow rates of reaction ingredients for synthesis of particles of different polydispersity indices (PDIs)}

        \begin{table}[htbp]
        \centering
\caption{The flow rates of the solutions of reaction ingredients during synthesis of PNIPAM particles of controlled PDIs.}
\vskip +0.1in

\begin{tabular}{|c|c|}
\hline
 Flow rate (ml/min)& PDI \% \\   \hline

0.8 & 9 \\
0.6 & 14 \\
0.7 & 15 \\
0.5 & 30 \\
\hline
\end{tabular}

\end{table}

\section{Lower critical solution temperature (LCST) of PNIPAM particles}
            PNIPAM particles are themoresponsive and their LCST in dilute aqueous suspension is estimated {\it via} differential calorimetry (DSC) and dynamic light scattering (DLS) experiments.  
            \begin{figure}[h]
	 	\includegraphics[width= 6.0in ]{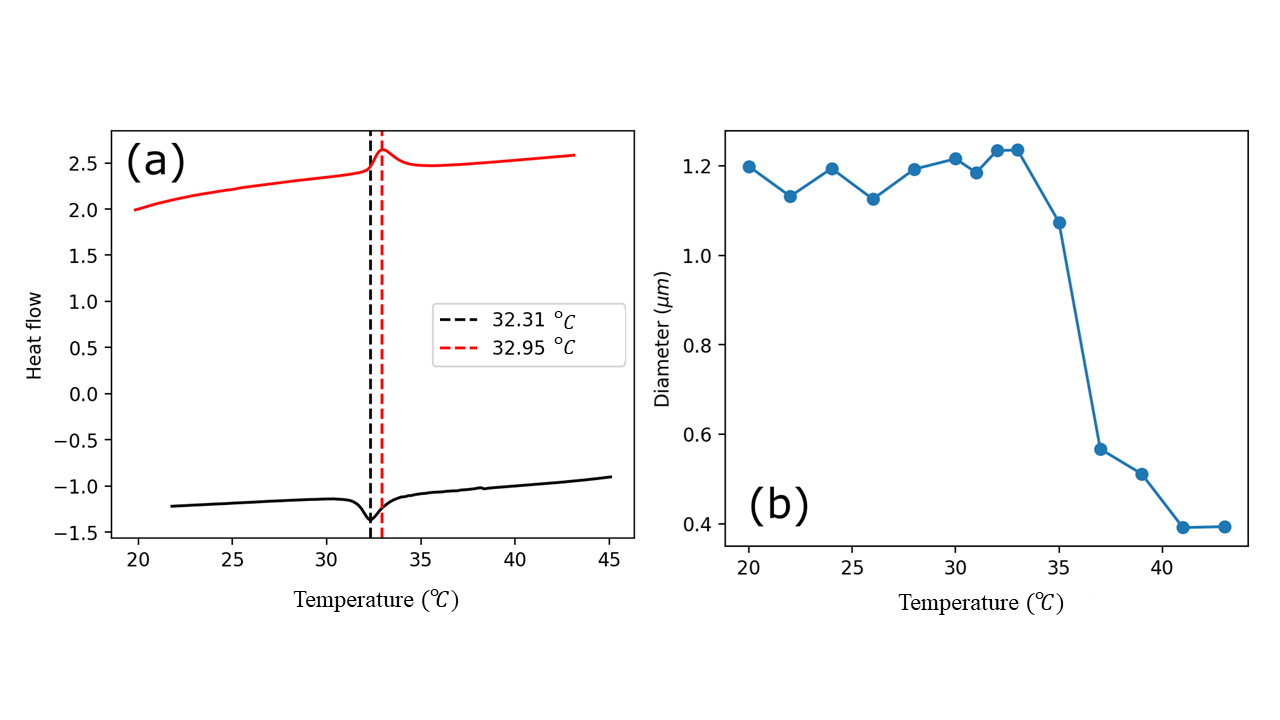}
	 	\centering
	 	\caption{\label{piezo} (a) Heat flow is measured using DSC after applying a temperature ramp to a PNIPAM suspension. The peak indicates an LCST of $\approx 32^\circ$ C. (b) Average hydrodynamic radii of PNIPAM particles are measured using DLS for different temperatures in dilute suspension. We see that particles deswell rapidly at LCST $\approx 32^\circ$ C. DSC and DLS therefore provide very consistent estimates of LCST.}
	       \end{figure}

         \section{Sample cell for confocal experiments}
            \begin{figure}[H]
            \centering
	 	\includegraphics[width= 5.0in ]{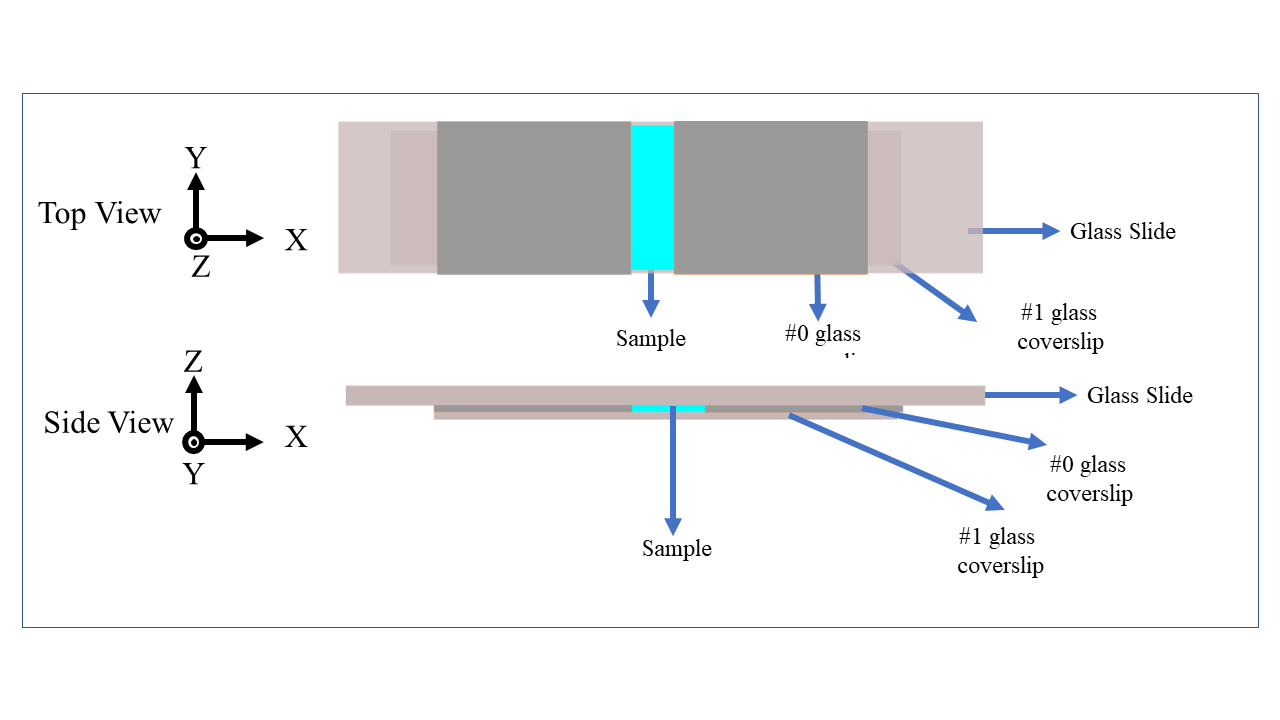}
            
	 	\caption{\label{cell} Sample cell for the confocal experiments, assembled using a glass slide, a \#1 coverslip, and two \#0 coverslips used as spacers.}
	       \end{figure}

    \section{Fitted parameters for Gaussian fit in Fig.1}      
    \begin{table}[htbp]
    
    \centering
    \caption{Fitted Parameters}
    
    \label{tab:fitted_parameters}
    \begin{tabular}{|c|c|c|c|c|c|}
    \hline
    Dataset & Amplitude & Mean ($M$) & Standard deviation ($W$)  & Offset & PDI(\%) \\
    \hline
    1 & 188.01 & 0.82 & 0.07 & 20.54 & 9 \\
    2 & 113.28 & 1.47 & 0.20 & 20.13 & 14 \\
    3 & 100.60 & 1.65 & 0.24 & 16.52 & 15 \\
    4 & 106.87 & 1.00 & 0.29 & 20.80 & 30 \\
    \hline
    \end{tabular}
   
    \end{table}

            \section{Dependence of $Q_2$ and $\chi_4$ on probe length $a$}
            \begin{figure}[H]
	 	\includegraphics[width= 6.0in ]{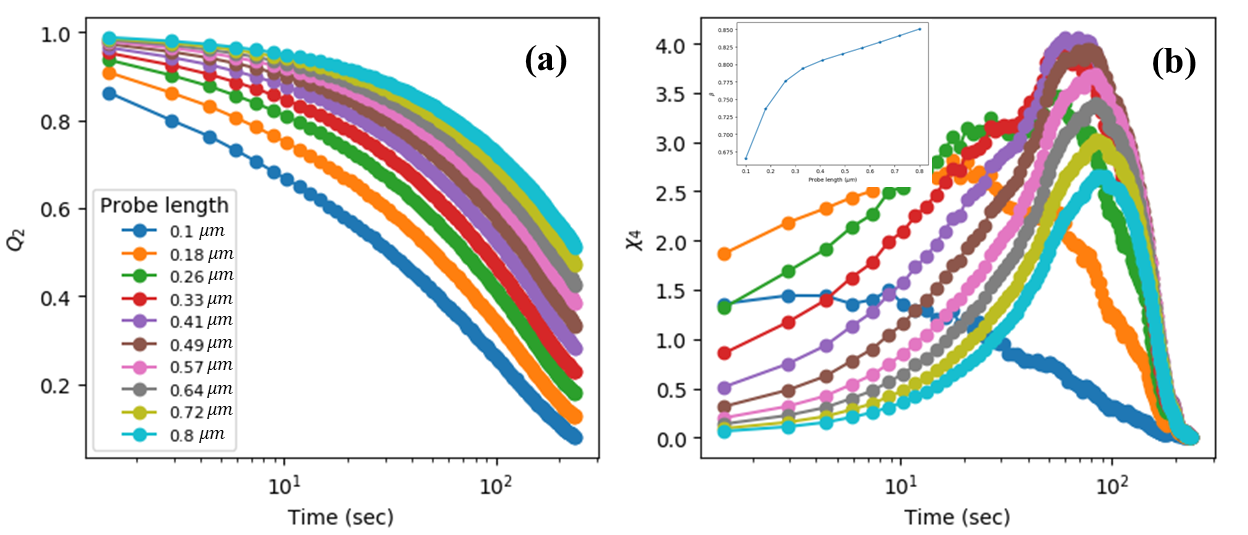}
	 	\centering
	 	\caption{ (a) $Q_2$, plotted for different probe lengths $a$ (shown in the legend), for aqueous suspensions of PNIPAM particles of PDI 9\%, prepared at an area fraction of 0.65. (b) The corresponding $\chi_4$ values at the same probe lengths (designated by the same colours as in S3(a)) are plotted vs time. In the inset, $\beta$ exponent values from the fits of $Q_2$ shown in (a) are plotted versus probe length.}
	       \end{figure}

       \section{Optimal probe lengths computed from experiments (2D) and simulations (3D)}  
       \begin{table}[htbp]
\centering
\caption{The optimal probe lengths, $a^*$, used for analyses of both experiments and simulations, are tabulated for suspensions of different PDIs and packing fractions. M is the mean obtained from the Guassian fit.}
\vspace{0.1cm}

\begin{tabular}{cccc|cccc}
\hline
 &\multicolumn{1}{c}{$Experimental$} &\multicolumn{6}{c}{$Simulation$}\\
 \hline
 PDI & $A_p$ & $a^*$ ($\mu$m) & $a^*$/M & PDI & $\phi$ & $a^*$  & $a^*$/$\sigma_{mean}$ \\   \hline
  9\% & 0.64 & 0.41& 0.50 & 13.06\% & 0.623 & 0.28 & 0.28\\ 
  9\% & 0.59 & 0.49& 0.60 & 17.37\% & 0.623 & 0.32 & 0.32\\ 
  9\% & 0.53 & 0.26& 0.31 & 21.77\% & 0.623 & 0.33 & 0.33\\ 
  14\% & 0.66 & 0.72& 0.49 & 38.67\% & 0.623 & 0.38 & 0.38\\ 
  15\% & 0.65 & 0.57& 0.34  & &  &  \\ 
  30\% & 0.65 & 0.57& 0.57 &  &  &  \\ 
  \hline
  
\end{tabular}
\end{table}

            \section{Experimental MSD and $\chi_4$ values, calculated from experiments at different temperatures of the suspension medium}
            
            \begin{figure}[H]
	 	\includegraphics[width= 6.0in ]{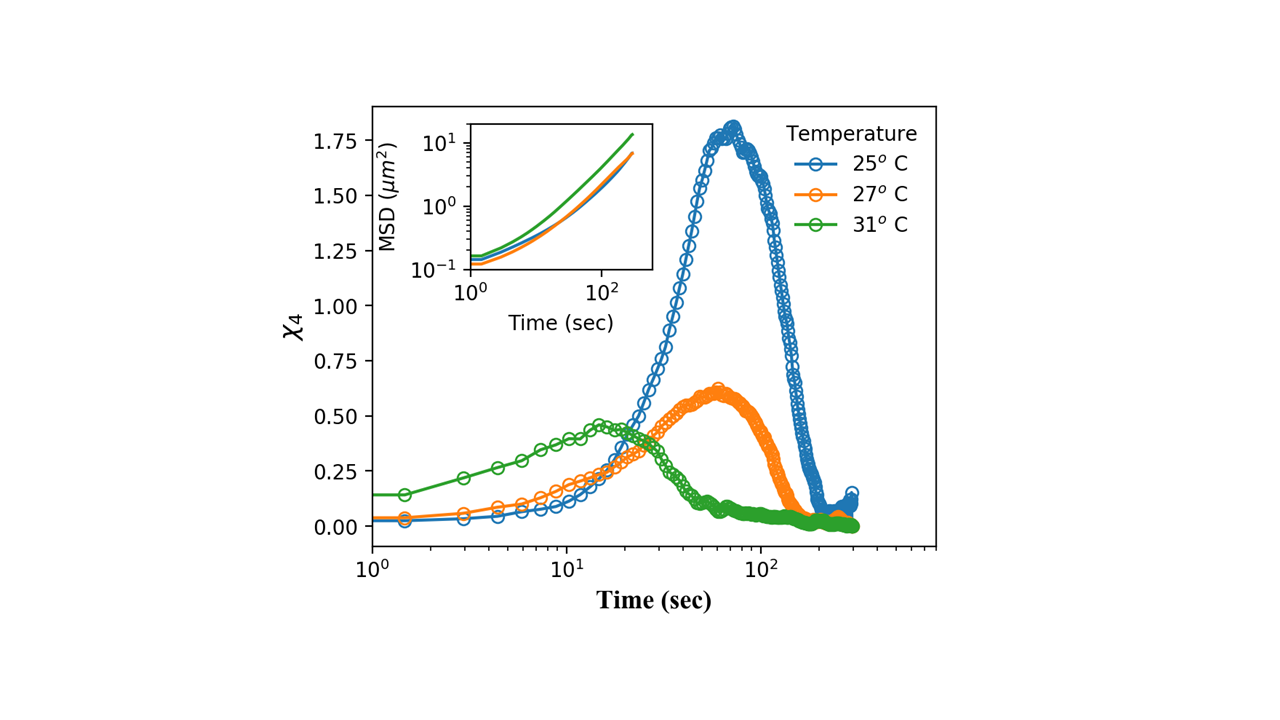}
	 	\centering
	 	\caption{ $\chi_4$ peak of a PNIPAM suspension of area fraction $A_p$= 0.65, constituted by particles of PDI = 14\%, decreases when temperature of the suspension medium is increased. Simultaneously, the inset shows an increase in the particle MSDs  with increase in temperature due to accelerated particle diffusion.}
	       \end{figure}

            \section{ MSD and $\chi_4$ values calculated from MD simulations by varying temperature and volume fraction}
            
            \begin{figure}[H]
            \centering
	 	\includegraphics[width= 5.0in ]{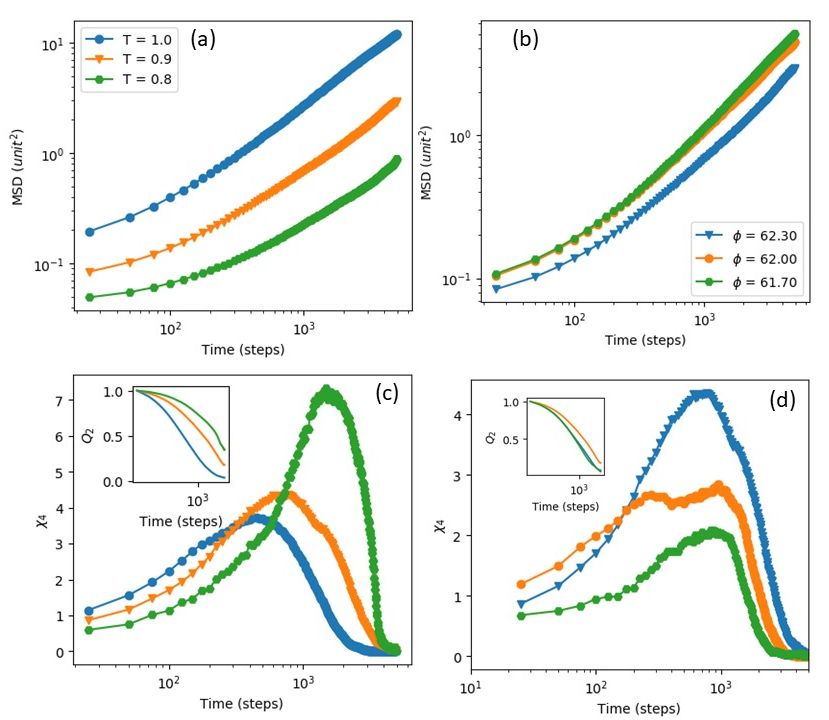}
	 	\caption{\label{QPD} (a) MSDs extracted from MD simulations are plotted $vs.$ time steps for three different temperatures at volume fraction $\phi=$ 0.6230. (b) MSDs $vs.$ time steps for different volume fractions at T=0.9. (c) $\chi_4$ values for the same temperatures and $\phi$ as in (a) for $a^* = 0.31, 0.28, 0.25$ for T = 1.0, 0.9 and 0.8 respectively. The corresponding $Q_2$ are shown in the inset. (d) $\chi_4$ for the same volume fractions as in (b) for T=0.9 with corresponding $a^* = 0.28, 0.32, 0.33$. The corresponding $Q_2$ are shown in the inset.}
	       \end{figure}

	
    \vspace{1 cm}